\documentclass[11pt]{article}
\usepackage{amsmath,amssymb,color,epsf}
\usepackage{amssymb,slashed,latexsym,amsmath,multirow,color,dsfont}
\usepackage{graphicx}
\usepackage{cite}
\usepackage{float}
\usepackage{epsfig}
\usepackage{soul}
\usepackage{color}

\makeatletter
\@addtoreset{equation}{section}
\makeatother

\usepackage{amsfonts}

\usepackage{amsfonts}

\usepackage[font=footnotesize,labelsep=newline,labelfont=sc,justification=centering,position=top]{caption}

\newcommand{\bea}{\setlength\arraycolsep{2pt} \begin{eqnarray}}
\newcommand{\eea}{\end{eqnarray}}

\usepackage{hyperref}

\newsavebox{\uuunit}
\sbox{\uuunit}
    {\setlength{\unitlength}{0.825em}
     \begin{picture}(0.6,0.7)
        \thinlines
        \put(0,0){\line(1,0){0.5}}
        \put(0.15,0){\line(0,1){0.7}}
        \put(0.35,0){\line(0,1){0.8}}
       \multiput(0.3,0.8)(-0.04,-0.02){12}{\rule{0.5pt}{0.5pt}}
     \end {picture}}

\def\be{\begin{equation}}
\def\ee{\end{equation}}
\def\ba{\begin{array}}
\def\ea{\end{array}}
\def\bea{\begin{eqnarray}}
\def\eea{\end{eqnarray}}
\def\bd{\begin{displaymath}}
\def\ed{\end{displaymath}}

\begin{document}

\begin{titlepage}

\begin{center}

\hfill UG-16-73 \\  

\vskip 1.5cm

{\Large \bf  Holographic P-wave Superconductors in 1+1 Dimensions}
\vskip 1cm

{\bf Gokhan Alkac\,$^{1, a}$,  Shankhadeep Chakrabortty\,$^{2, b}$, Pankaj Chaturvedi\,$^{3, c}$. } \\

\vskip 25pt

{\em $^1$ \hskip -.1truecm Van Swinderen Institute for Particle Physics and Gravity,  \\ 
University of Groningen, Nijenborgh 4, 9747 AG Groningen, The Netherlands \vskip 5pt }

{\em $^2$ Department of Physics, Indian Institute of Technology Ropar, Rupnagar, \\ Punjab 140001, India.}

{\em $^3$ Yau Mathematical Sciences Center Tsinghua University, Beijing, 100084, China}

{email: {\em $^{a}$ g.alkac@rug.nl, \em $^b$ s.chakrabortty@iitrpr.ac.in, \em $^c$ cpankaj1@gmail.com} } \\

\vskip 25pt

\end{center}

\vskip 0.5cm

\begin{center} {\bf ABSTRACT}\\[3ex]
\end{center}

We study $(1+1)$-dimensional p-wave holographic superconductors described by three dimensional Einstein-Maxwell gravity coupled to a massive complex vector field in the context of $AdS_3/CFT_2$ correspondence. In the probe limit where the backreaction of matter fields is neglected, we show that there is a formation of a vector hair around the black hole below a certain critical temperature. In the dual strongly coupled $(1+1)$-dimensional  boundary theory, this holographically corresponds to the formation of a charged vector condensate which breaks spontaneously both the $U(1)$ and $SO(1,1)$ symmetries. We numerically compute both the free energy and the ac conductivity for the superconducting phase of the boundary field theory. Our numerical computations clearly establish that the superconducting phase of the boundary theory is favorable over the normal phase and the presence of a magnetic moment term in the dual bulk theory effects the conductivity in the boundary field theory.

\end{titlepage}

\tableofcontents

\setcounter{page}{1}

\section{Introduction}
The {\it gauge/gravity correspondence} stands as one of the most significant developments in the theoretical physics over the last decade. In the large N limit, this duality relates  a weakly coupled theory of gravity in the $AdS$ spacetime to a strongly coupled conformal field theory living on the  boundary of the $AdS$ spacetime  \cite{Maldacena:1997re,Gubser:1998bc,Witten:1998qj,Aharony:1999ti}. In the subsequent studies, it is also shown that the strongly coupled boundary conformal field theories at finite temperature can be holographically related to the bulk geometries described by the $AdS$-black hole \cite{Witten:1998zw}. In this setup, the Hawking temperature of the black hole  is then considered to be the temperature of the strongly coupled boundary field theory. Moreover, the Hawking-Page phase transition for the $AdS$-black hole in the bulk holographically captures the confinement-deconfinement phase transition in the boundary field theory. Later, it is realized that the strongly coupled boundary field theories at a finite temperature and as well as a finite charge density can also be considered to be the holographic dual to a gravitational theory coupled to the gauge fields and other matter fields, admitting a charged $AdS$ black hole solution. This form of holographic duality endows us an extremely powerful tool for describing strongly coupled condensed matter systems at finite temperature and chemical potential. In this context, the study of superconductivity in strongly coupled condensed matter systems has been initiated with the Gubser's seminal work \cite{Gubser:2008px}, where the author shows that a charged scalar living on Reissner-Nordstrom-AdS black hole background becomes unstable enough near the black hole horizon to form a condensate which can mimic the behavior of electrons forming the Cooper pair in a superconductor. Later, it has been shown by Hartnoll et. al. \cite{Hartnoll:2008vx,Hartnoll:2008kx} that this phase transition in the bulk as a result of the instability of the matter field (a charged scalar), corresponds to a superconducting phase transition in the dual strongly coupled boundary field theory. The boundary field theory mentioned above describes an s-wave superconductor signifying the zero orbital angular momentum of the corresponding Cooper pair wave function and it is dual to an Abelian Higgs model with a bulk complex scalar field charged under the Maxwell field \cite{Horowitz:2008bn,Herzog:2009xv,Horowitz:2010gk,Herzog:2010vz,Sachdev:2010ch}.

Although the conventional superconductors can be very well described by BCS theory, a complete understanding of the high-$T_c$ superconductors, especially the pairing mechanism still remains far to be achieved till date. It is commonly believed that the theory of  high-$T_c$ superconductors  cannot be tackled with standard perturbative methods, a tractable strongly coupled theory might provide insight to this problem. In particular, being the most promising candidate to explore strongly coupled non-perturbative systems, $gauge/gravity$ duality provides us an alternative way to study the high-$T_c$ superconductors. Following the holographic prescription, in all the pioneering works previously mentioned,  an Abelian Higgs model with a charged scalar field on a black hole background is employed in the bulk side of the correspondence. The notion of temperature is provided by the black hole and below a critical temperature the formation of a charged scalar hair is observed. This implies in the boundary field theory side of the correspondence that a charged scalar operator acquires a nonzero vacuum expectation value below the critical temperature, breaking the global U(1) symmetry. The two main electronic properties of a superconductor, namely a divergent Drude peak at zero frequency and the formation of a gap in the real part of the electric conductivity, are also achieved. Diverging conductivity is hardly surprising because in a normal conductor the breaking of the translational invariance ensures a finite value of the conductivity at any frequency whereas the boundary field theories with a superconducting phases are fully translationally invariant. Moreover, the transition between the normal and the superconducting phase is studied both inside the probe limit  by neglecting the backreaction of matter fields to the background geometry\cite{Hartnoll:2008vx}, and by explicitly taking the backreaction into account \cite{Hartnoll:2008kx}. In the latter case, one observes a harder condensation, i.e. the transition temperature is shifted to lower values.

In a real superconductor, due to the attractive interaction mediated by phonons, there is a formation of the Cooper pair consisting of electrons with opposite spins and momenta. The Cooper pair wave function plays the role of order parameter in the phase transition and its orbital angular momentum $\ell$ can take values as  $\ell = 0$ (s-wave), $\ell = 1$ (p-wave) and  $\ell = 2 $ (d-wave) etc. While an s-wave holographic superconductor can be studied with the use of a charged scalar field coupled to Einstein-Maxwell system as mentioned before, d-wave superconductors can be realized holographically by the condensation of a charged spin-2 matter field in a similar fashion \cite{Kim:2013oba,Benini:2010pr,Chen:2010mk}. For p-wave superconductors, various models have been proposed. The first model makes use of an SU(2) Yang-Mills field in the bulk and one of the gauge degrees of freedom is considered to be dual to spin-1 order parameter in the field theory side of the correspondence \cite{Gubser:2008wv}. In the two other models, superconductivity is achieved by the condensation of various charged matter fields in the bulk, namely a 2-form field \cite{Donos:2011ff,Donos:2012gg} and a massive spin-1 vector \cite{Cai:2013pda,Cai:2013aca}. In the case of charged, massive spin-1 condensate, it is shown that the magnetic moment of the spin-1 field plays a crucial role when the phase transition is induced by an external magnetic field \cite{Cai:2013pda}. Other important studies along this line of research can be found in \cite{Gregory:2009fj,Liu:2010ka,Gubser:2008pf,Horowitz:2010jq,Brihaye:2010mr,Domenech:2010nf,Roberts:2008ns,Horowitz:2009ij,Amado:2009ts,Nishioka:2009zj,Basu:2009vv,Bu:2012zz,Herzog:2009ci,Ammon:2008fc,Ammon:2009fe,Peeters:2009sr,Benini:2010qc,Sonner:2009fk,Franco:2009yz,Siopsis:2010uq}.

The study of (1+1)-dimensional s-wave holographic superconductors was initiated in \cite{Ren:2010ha}, followed by further generalizations and extensions in \cite{Liu:2011fy,Nurmagambetov:2011yt,Zeng:2012xy,Momeni:2013waa,Momeni:2013eva,Chaturvedi:2013ova,Peng:2015lsa,Peng:2016mxd}. (1+1)-dimensional p-wave holographic superconductors are investigated using the probe brane construction in \cite{Bu:2012qr}. Moreover, there is  further improvements of holographic realization for p-wave superconductor using a model described as an Einstein-Maxwell system coupled to a charged, massive spin-1 matter field \cite{Cai:2013pda,Cai:2013aca}. In this present work, we focus on the same Einstein-Maxwell-charged vector model of p-wave holographic superconductor in the context of AdS$_3$/CFT$_2$ correspondence. Although this model has been already examined in the four-dimensional bulk, it is by no means trivial to study the same model in three dimensions. Apart from various technical differences appearing in the formation of the condensate and the computation of conductivity of the dual field theory, we present a remarkable feature of this model which exhibits itself only in the case of three dimensional gravitational bulk but not in the higher dimensions. \textit{It turns out that in our present analysis, the ac conductivity of the dual field theory depends on the magnetic moment of the spin-1 field in the absence of an external magnetic field.} On the contrary, in higher dimensional cases, the magnetic moment can play a role in the physics of this model only when an external magnetic field acts on the system. This peculiar feature of the ac conductivity in (1+1) dimensional system within the regime holographic duality is the main motivation for the present work. 

The organization of the paper is as follows. In Section 2, we introduce the bulk gravitational theory that we use and give the field equations. Section 3 is devoted to the details of normal and superconducting phases. We reserve section 4 in order to compute the free energy of the system for both normal and superconducting phases. In section 5, we give details of the computation of conductivity for the dual field theory and compare it with the higher dimensional cases. In Section 6, we conclude with a summary of our results, discussions and some possible future investigations of this model.

\section{Model}
In this work, we use a bulk model in (2+1) dimensions dual to a (1+1) dimensional boundary field theory with a superconducting phase. The main feature of the model captures a spontaneous breaking of local U(1) symmetry in the bulk that holographically corresponds to a spontaneous breaking of global U(1) symmetry at the boundary. It is described by the Einstein-Maxwell action with a cosmological constant, coupled to a charged, massive spin-1 field as follows \cite{Cai:2013pda,Cai:2013aca}
\begin{equation}
\label{model}
S = \frac{1}{2 \kappa^2} \int d^3x \, \sqrt{-g} \left( R+\frac{2}{L^2}-\frac{1}{4}F_{\mu\nu} F^{\mu\nu}- \frac{1}{2} \rho_{\mu\nu}^\dagger\rho^{\mu\nu}-m^2 \rho_\mu^\dagger \rho^\mu + iq\gamma\rho_\mu \rho_\nu^\dagger F^{\mu\nu}\right),
\end{equation}
where $L$ is the AdS length and $\kappa^2=8\pi G$ characterizes the Newton's constant in the bulk. Moreover,  $q$ and $m$ are the electric charge and the mass of the spin-1 field respectively.  We employ the usual Maxwell field strength as $F_{\mu\nu}=2 \nabla_{[\mu}A_{\nu]}$ and $\rho_{\mu\nu}=2 D_{[\mu}\rho_{\nu]}$ with $D_\mu =\nabla_{\mu} - i q A_\mu$. The last term in the bulk action (\ref{model}) represents a nonlinear interaction between the complex vector field $\rho_\mu$ and the gauge field $A_\mu$ with $\gamma$ being the magnetic moment of the vector field $\rho_\mu$. In \cite{Cai:2013pda}, it has been shown that this interaction term has a significant effect in the case of a 4-dimensional bulk theory when an external magnetic field is applied to the system. A similar study can be be performed to investigate this effect in the 3-dimensional bulk where the magnetic field is a pseudo-scalar. However, we will focus our attention to phase transition and conductivity of boundary field theory in the absence of any external magnetic effect. As a nice example of richness of lower dimensional physics, this nonlinear term will have a nontrivial role in the conductivity properties of the boundary field theory.

We consider this model in the probe limit by neglecting the backreaction of the matter fields to the background geometry. While this is in general enough to probe the main characteristics of the phase transition and conductivity properties, it is very well known that further study by including backreaction effects can be useful especially for a better understanding of the phase transition \cite{Hartnoll:2008kx}. In order to work in the probe limit, it is customary to scale the fields as $\rho_\mu \rightarrow \rho_\mu/q$, $A_\mu \rightarrow A_\mu/q$ and take the limit $q\rightarrow\infty$. Then, the gravitational field equations become source-less Einstein's equations with the cosmological constant and the matter equations read as
\begin{eqnarray}
\nabla^\nu F_{\nu\mu} &=& i(\rho^\nu \rho_{\nu\mu}^\dagger - \rho^{\nu\dagger} \rho_{\nu\mu}) + iq\gamma \nabla^\nu(\rho_\nu \rho_\mu^\dagger - \rho_\nu^\dagger \rho_\mu), \label{Maxwell}\\ 
D^\nu \rho_{\nu\mu} &=& m^2 \rho_\mu - iq\gamma \rho^\nu F_{\nu\mu} \label{rho}.
\end{eqnarray}
The first one among the above set of equations is the Maxwell's equation sourced by the charged spin-1 field and the second one is the field equation for the spin-1 field itself. 
In the boundary theory,  the vacuum expectation value of a charged vector operator plays the role of order parameter in the theory. According to the standard AdS/CFT dictionary, this charged vector operator is sourced by the boundary value of the bulk field $\rho_{\mu}.$   In the bulk side, as first demonstrated in the s-wave case \cite{Hartnoll:2008vx,Hartnoll:2008kx}, the charged matter field becomes tachyonic enough when the temperature is lower than a critical value $T_c$ to develop a nontrivial profile around the black hole. The formation of this vector hair in the bulk side corresponds to a superconducting phase transition as the dual charged operator acquires a nonzero vacuum expectation value below the critical temperature, breaking the U(1) and SO(1,1) symmetries at the boundary.  The main characteristic of p-wave superconductors in the case of the 4-dimensional bulk theory is that  the condensation breaks the SO(2) rotational symmetry at the boundary, creating an anisotropy in spatial coordinates. However, in our holographic set up this cannot hold since only one spatial dimension is available at the boundary. 

With this brief description of three dimensional holographic model realizing a  superconducting phase transition in the boundary theory, in the next section, we elaborate more on how to attain the various phases  as well as transition among those phases.

\section{Phase transition: Normal and  Superconducting Phases}
To understand the superconducting phase of the strongly coupled (1+1) boundary theory, we explore further its holographic dual described by a (2+1) dimensional gravity theory coupled to Maxwell field as well as a charged spin-1 field.  As mentioned previously, the spin-1 field develops a nontrivial profile in the bulk below a certain temperature, resulting in  two distinct phases in the gravity side of the correspondence. Accordingly, the holographic correspondence suggests that the charged operator dual to the spin-1 bulk field acquires a nonzero vacuum expectation value at the boundary field theory, and becomes the order parameter distinguishing the normal and superconducting phases. It is important to note that above the critical temperature $T_c$, the gravity theory admits a charged BTZ black hole solution that holographically corresponds to the normal phase of the boundary theory.  Whereas, below $T_c$, the gravity theory allows a non-vanishing profile of a charged spin-1 field around the charged BTZ black hole, signifying the superconducting phase of the boundary theory. In this present section, we give the details of the holographic description of both phases of the boundary theory and the corresponding phase transition among them.

\subsection{Normal Phase}
Above the critical temperature $T_c$, the bulk description of the normal phase is characterized by the vanishing of the spin-1 field $\rho_\mu=0$ and the bulk model that we use here reduces to the Einstein-Maxwell theory with cosmological constant in three dimensions. In this phase, the allowed solution is the charged BTZ black hole described by 
\begin{eqnarray}
ds^2 &=& -f(r)dt^2+\frac{dr^2}{f(r)}+r^2 dx^2, \\ 
f(r) &=& r^2-r_h^2+\frac{\mu^2}{2 r^2}\ln\left(\frac{r_h}{r}\right),\\
A_\mu dx^\mu &=& \mu \ln \left(\frac{r_h}{r}\right),
\end{eqnarray}
where $\mu$ is the chemical potential,  $r=r_h=1$ corresponds to the horizon radius and the Hawking temperature of the black hole is given by $T_{Bh}=\frac{\mid f'(r_h)\mid}{4 \pi}=\frac{\mid 4-\mu^2\mid}{8 \pi}$ . In what follows we show the formation of a vector hair below the critical temperature $T_c$, giving rise to the superconducting phase in the dual field theory.

\subsection{Superconducting Phase}
In order to show the existence of the superconducting phase, we work within the probe limit and use the BTZ black hole as the background spacetime to solve equations for the gauge field and the spin-1 field. By adopting the following ansatz 
\begin{equation}
A_\mu dx^\mu = \phi(r) dt,\quad \rho_\mu dx^\mu = \rho_x(r) dx,
\end{equation}
we obtain the equations for the electric field $\phi$ and the x-component of the spin-1 field $\rho_x$ as
\begin{eqnarray}
\phi^{\prime\prime} + \frac{1}{r}\phi^\prime - \frac{2  \rho_x^2}{f r^2} &=& 0 \label{phieqn}  \\
\rho_x^{\prime\prime} + \left(\frac{f^\prime}{f}-\frac{1}{f}\right)\rho_x^\prime + \left(\frac{\phi^2}{f^2}-\frac{m^2}{f}\right) \rho_x &=& 0. \label{rhoxeqn}
\end{eqnarray} 
Taking the boundary limit of these equations ($r \rightarrow \infty$), one can find the boundary behavior of the fields as
\begin{equation}
\phi = -\mu \log(r) + \rho + \ldots,\;\;\;\;\;\;
\rho_x = \frac{\rho_{x-} }{r^{\Delta_-}} + \frac{\rho_{x+} }{r^{\Delta_+}} + \ldots,\;\;\;\;\;\;\Delta_{\pm} = \pm m. \label{boundary}
\end{equation}
where we interpret $\mu$, $\rho$, $\rho_{x-}$, $\rho_{x+}$ as the chemical potential, the charge density, the source and the vacuum expectation values of the operator coupling to $\rho_x$ at the boundary field theory respectively.\footnote{From this computation, it can also be seen that the BF-bound \cite{Breitenlohner:1982bm} for a spin-1 particle in three dimensions is $m^2>0.$} $\rho_{x-}$ should be set to zero as a boundary condition to ensure that the phase transition is realized in the absence of an external source. 
\begin{figure}[H]
\centering
\begin{minipage}[b]{.7\linewidth}
\includegraphics[width =4in,height=2.8in]{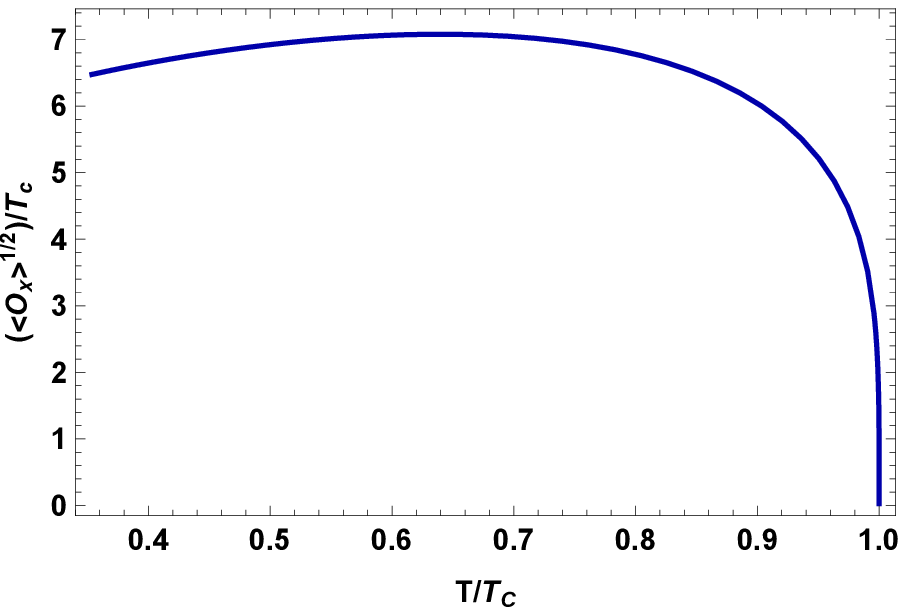}
\end{minipage}
\caption{\label{fig:Condensate} Condensation of the vector operator for $m^2=1/4$.} \end{figure}   
We impose $\phi(1)=0$ such that the norm of the gauge field $A_\mu$ remains finite at the horizon and the x-component of the spin-1 field satisfies
\begin{equation}
\rho_x(1)=\frac{2}{m^2}\rho_x'(1).
\end{equation}
We solve the equations for the electric field $\phi$ and the x-component of the spin-1 field $\rho_x$ (\ref{phieqn})-(\ref{rhoxeqn}) numerically subjected to the boundary conditions both at the horizon and the boundary for a particular choice of mass, $m^2 = \frac{1}{4}$, satisfying the BF-bound \cite{Breitenlohner:1982bm} for the system we consider here. As shown in the figure (\ref{fig:Condensate}), the condensation occurs below the critical temperature $T_c$ and the curve is qualitatively similar to what one expects from BCS theory.  However, it is important to note that in our case the condensate starts decreasing slightly starting from $T/T_c=0.6$. This might be due to the fact that we work within the probe limit approximation. Nevertheless, to show the condensation of the operator this approximation turns out to be legitimate.  In the next section, we will study the free energy difference of the normal and superconducting phases and determine which phase is thermodynamically favorable below the critical temperature $T_c$.
 
\section{Free Energy}
In this section, we calculate the free energy of the system for both normal and superconducting phases by working in the canonical ensemble where the charge density is fixed. The free energy is holographically identified with temperature times the on-shell Euclidean action. Since we work in the probe limit, neglecting the gravitational part, we write the Euclidean action as
\begin{equation}
-2\kappa^2 S_E=\int d^3x \, \sqrt{-g}\frac{1}{2}A_\nu \nabla_\mu F^{\mu\nu}-\int d^2x \, \sqrt{-h} n_\mu \Big(\frac{1}{2}A_\nu F^{\mu\nu}+\rho_{\mu}^\dagger \rho^{\mu\nu}\Big)+S_{ct}^{(\rho)}+S_{ct}^{(A)}, \label{Euc}
\end{equation}
by employing field equations (\ref{Maxwell}-\ref{rho}). $h_{\mu\nu}$ is the induced metric at the boundary ($r\rightarrow \infty$) and $h$ denotes its determinant. Two counter terms we introduce here to make the action finite are given by
\begin{eqnarray}
S_{ct}^{(\rho)}&=& -\Delta_-\int d^2x \, \sqrt{-h} h^{\mu\nu}\rho_{\mu}^\dagger \rho_{\nu}, \nonumber \\
S_{ct}^{(A)}&=&\frac{1}{2 \ln \Lambda}\int d^2x \, \sqrt{-h} h^{\mu\nu}A_{\mu} A_{\nu}.
\end{eqnarray}

\begin{figure}[H]
\centering
\begin{minipage}[b]{.7\linewidth}
\includegraphics[width =3.5in,height=2.5in]{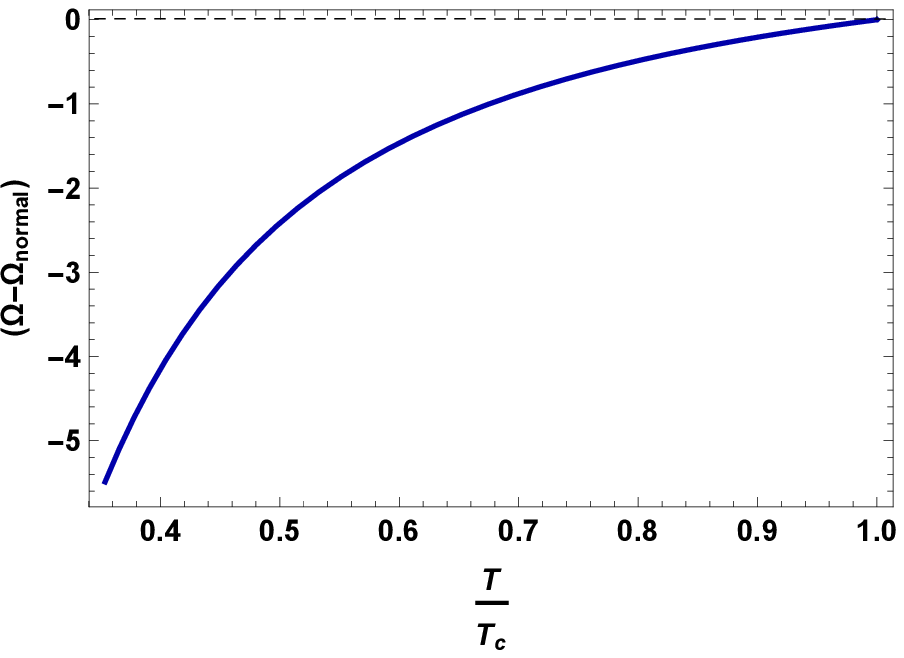}
\end{minipage}
\caption{\label{fig:FreeEn1by4} Plot showing the free energy difference of the superconducting and the normal phase ($\Omega-\Omega_{normal}$) verses temperature for $m^2=1/4$.} \end{figure}   

The first term is required to remove the infinities arising from the boundary expansion of massive spin-1 field $\rho_{\mu}$. The second term is needed to regularize the logarithmic divergence which appears in the boundary expansion of the gauge field $A_\mu$. It is not needed in higher dimensions and requires special care. The boundary integrals containing the gauge field $A_\mu$ should be evaluated at $r=\Lambda$ and then one takes the limit $\Lambda \rightarrow \infty$ at the end (see \cite{Hung:2009qk} for details). Substituting the boundary behavior of fields (\ref{boundary}) into the Euclidean action (\ref{Euc}) gives the free energy as
\begin{equation}
\Omega=\frac{2\kappa^2 F}{V}=\frac{1}{2}\mu\rho - 2\Delta_+\,\rho_{x-}\rho_{x+}-\int_{r_h}^{\infty}dr\,  \sqrt{-g}\frac{1}{2}A_\nu \nabla_\mu F^{\mu\nu},
\end{equation}
where $V=\int dx$. Using this expression, the free energy difference between the superconducting and the normal phase denoted by ($\Omega-\Omega_{normal}$) can be computed as a function of temperature. As can be seen from Figure \ref{fig:FreeEn1by4}, the superconducting phase is thermodynamically favored below the critical temperature $T_c$. The phase transition is second order, which can be checked from the derivative of free energy with respect to temperature. We now move to the next section in order to compute the conductivity in the superconducting phase using the holographic dual bulk model.

\section{Conductivity}
In this section, we compute the conductivity of the boundary field theory by introducing appropriate electromagnetic perturbation on the black hole background. This electromagnetic perturbation in the bulk results in a current at the boundary which is sourced by the boundary value of the gauge field. For a (1+1)-dimensional field theory, the conductivity is defined as
\begin{equation}
J_x=\sigma_{xx} E_x,
\end{equation}
not allowing Hall conductivity since there is only one spatial dimension.

Incidentally,  to introduce the electromagnetic perturbation in a consistent way one needs to  turn on also matter perturbations as follows\footnote{This also occurs in some models of d-wave \cite{Benini:2010pr} and $d + i d$ \cite{Chen:2011ny} holographic superconductors.}
\begin{eqnarray}
A_\mu dx^\mu &=& e^{-i \omega t} A_x dx \nonumber  \\ 
\rho_\mu dx^\mu &=& e^{-i \omega t} \rho_t(r) dt+e^{-i \omega t} \rho_r(r) dr, \nonumber  \\
\rho_\mu^{\dagger} dx^\mu &=& e^{-i \omega t} \rho_t^{\dagger}(r) dt+e^{-i \omega t} \rho_r^{\dagger}(r) dr,     
\end{eqnarray}
where the perturbations $\rho_\mu$ and $\rho_\mu^\dagger$ are treated independently. Inserting the above form of perturbations into the matter equations (\ref{Maxwell})-(\ref{rho}) results in the following linearized equations
\begin{eqnarray}
 0 &=& A_x^{\prime\prime} +  \left(\frac{f^\prime}{f}-\frac{1}{r}\right)A_x^\prime + \frac{\omega^2}{f^2}A_x
 + \frac{i \rho_x}{r f^2}  \Big(-i r (\rho_t(\gamma \omega - \phi) - \rho_t^*(\gamma \omega + \phi)+ \nonumber  \\ 
&& r \gamma f(-\rho_r+\rho_r^*)f^\prime + \gamma f^2(\rho_r - \rho_r^* - r \rho_r^\prime + r \rho_r^{\prime *})) 
 -i (1+\gamma)(\rho_r - \rho_r^*)\rho_x^\prime\Big) \\
 0 &=& \rho_t^{\prime \prime} + \frac{1}{r} \rho_t^{\prime} - \frac{m^2}{f} \rho_t + \frac{i \rho_x}{r f^2}   \Big( rf (r(\omega + \phi)\rho_r^\prime + \rho_r(\omega + \phi \ + r(1+\gamma)\phi^\prime) \nonumber \\
&&+ r \gamma f(-\rho_r+\rho_r^*)f^\prime + \gamma f^2(\rho_r - \rho_r^* - r \rho_r^\prime + r \rho_r^{\prime *})) i A_x \rho_x (\gamma \omega - \phi)\Big) \\
0 &=& -r^2 \rho_r(-m^2 f +(\omega + \phi)^2) + i r^2(\omega + \phi)\rho_t^{\prime} - i \Big(f(-\gamma \rho_x A_x^\prime + A_x \rho_x^\prime) \nonumber \\
&&+ r^2 \gamma \rho_t \phi^\prime \Big).    
\end{eqnarray}

Moreover, the equations for $\rho_t^\dagger$ and $\rho_r^\dagger$  can be obtained by the set of replacements $\rho_t \rightarrow \rho_t^\dagger$, $\rho_r \rightarrow \rho_r^\dagger$ and $\omega \rightarrow -\omega$. Since the equations for $\rho_r$ and $\rho_r^\dagger$ are algebraic, it is straightforward to solve the corresponding equations for these radial components of the matter perturbation and utilize the result to obtain a set of second order coupled differential equations for $A_x$, $\rho_t$ and $\rho_t^\dagger$. Furthermore it is evident from the form of these set of coupled differential equations that the necessity of introducing matter perturbations plays the role of consistency conditions in the bulk theory. For example, without any matter perturbation, field equations for the spin-1 field forces to $A_x=0$. As shown in [20], in higher dimensional cases, one can introduce the electromagnetic perturbation in another spatial direction such as $A_y$ and the equation for this perturbation decouples from the rest of the perturbations. Therefore, one can compute conductivity from the equation for the perturbation $A_y$ alone and there is no dependence in the magnetic moment. In the case of three dimensional bulk, however, the consistency of $A_x$ perturbations forces us to introduce matter perturbations and as a result of that conductivity becomes dependent on the magnetic moment.

In order to solve the perturbation equations, one should impose the near horizon behavior of the fields as
\begin{eqnarray}
\label{fieldhor}
&& A_x=c_1(1-r^{-1})^{\pm i \omega /2},\quad \rho_t=c_2(1-r^{-1})^{\pm i \omega /2},\quad \rho_t^\dagger=c_3(1-r^{-1})^{\pm i \omega /2}, \nonumber \\
&& \rho_r=c_4(1-r^{-1})^{\pm i \omega /2-1},\quad \rho_r^\dagger=c_5(1-r^{-1})^{\pm i \omega /2-1},
\end{eqnarray}
where $c_i$'s are arbitrary constants. The last two equations in (\ref{fieldhor}) are the consequences of the fact that the spin-1 field $\rho_\mu$ has a finite norm at the horizon. We choose minus sign to impose the ingoing wave boundary condition which is required to ensure causality. Once the parameters $(c_1,c_2,c_3)$ are fixed, equations for $A_x$, $\rho_t$ and $\rho_t^\dagger$  can be integrated out to the boundary.

\begin{figure}[H]
\centering
\begin{minipage}[b]{0.5\linewidth}
\includegraphics[width =2.5in,height=1.6in]{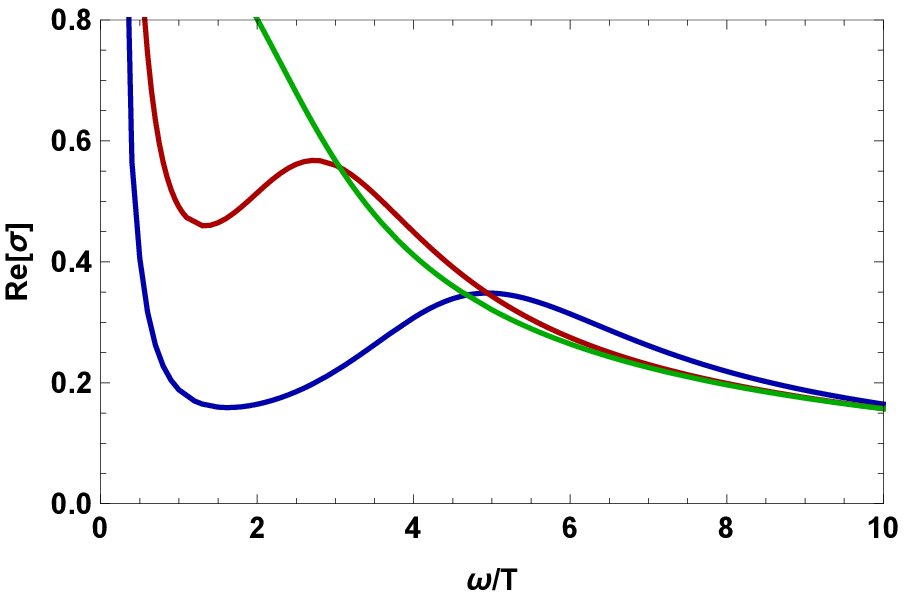}
\end{minipage}%
\begin{minipage}[b]{0.5\linewidth}
\includegraphics[width =2.5in,height=1.6in]{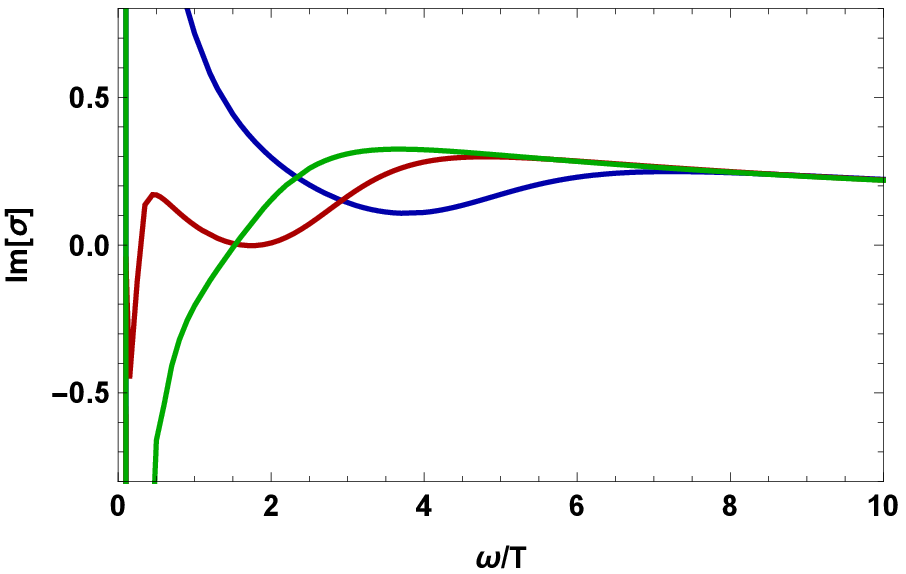}
\end{minipage}\quad
\caption{\label{fig:Conduct_0p01}For $m^2=1/4$, real and imaginary parts of the ac conductivity for superconducting phase are plotted with respect to $\omega/T$ for fixed value of $\gamma=0.01$. The blue, red and green curves correspond to the temperatures $T/T_c\approx0.4$ , $T/T_c\approx0.6$ and $T/T_c\approx0.8$ respectively. } \end{figure}

\begin{figure}[H]
\centering
\begin{minipage}[b]{0.5\linewidth}
\includegraphics[width =2.5in,height=1.6in]{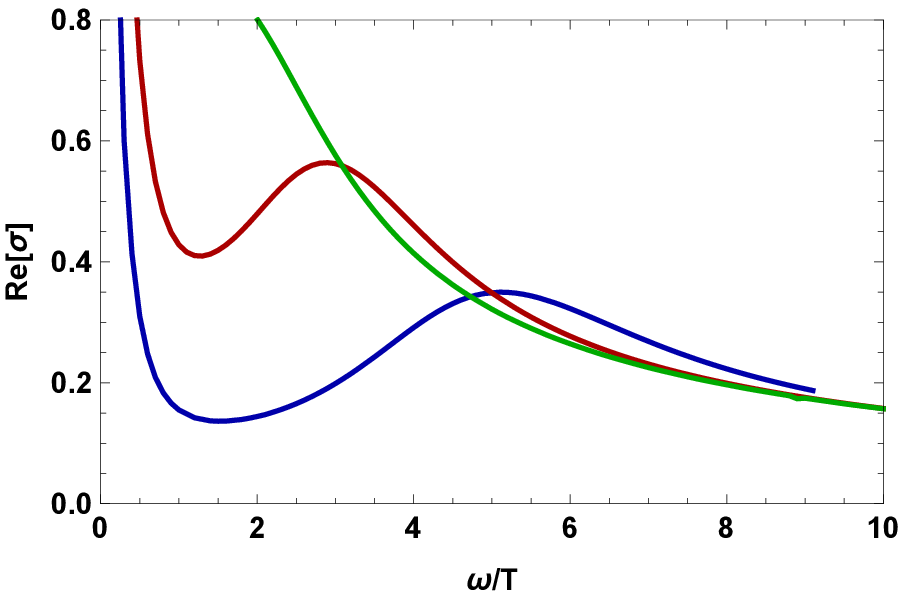}
\end{minipage}%
\begin{minipage}[b]{0.5\linewidth}
\includegraphics[width =2.5in,height=1.6in]{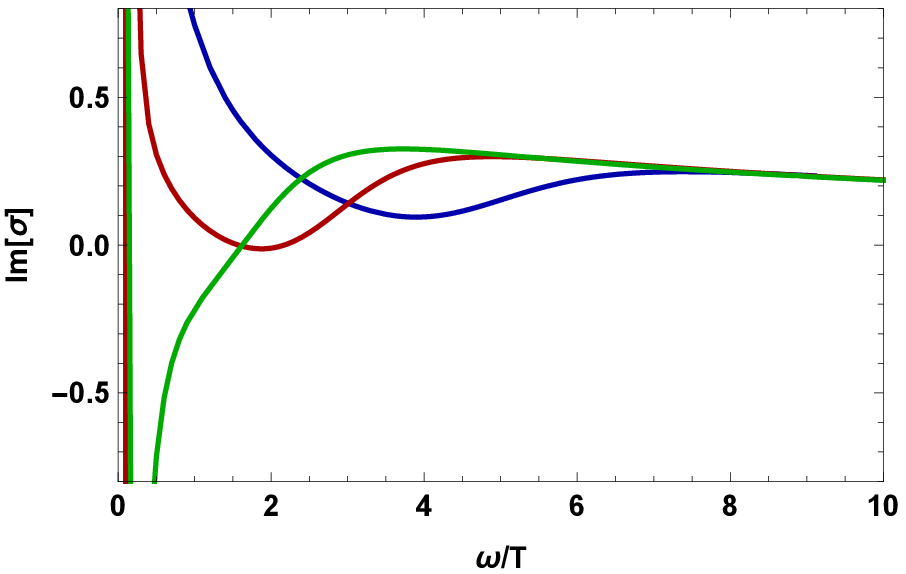}
\end{minipage}\quad
\caption{\label{fig:Conduct_0p1}For $m^2=1/4$, real and imaginary parts of the ac conductivity for superconducting phase are plotted with respect to $\omega/T$ for fixed value of $\gamma=0.1$. The blue, red and green curves correspond to the temperatures $T/T_c\approx0.4$ , $T/T_c\approx0.6$ and $T/T_c\approx0.8$ respectively.  } \end{figure}

\begin{figure}[H]
\centering
\begin{minipage}[b]{0.5\linewidth}
\includegraphics[width =2.5in,height=1.6in]{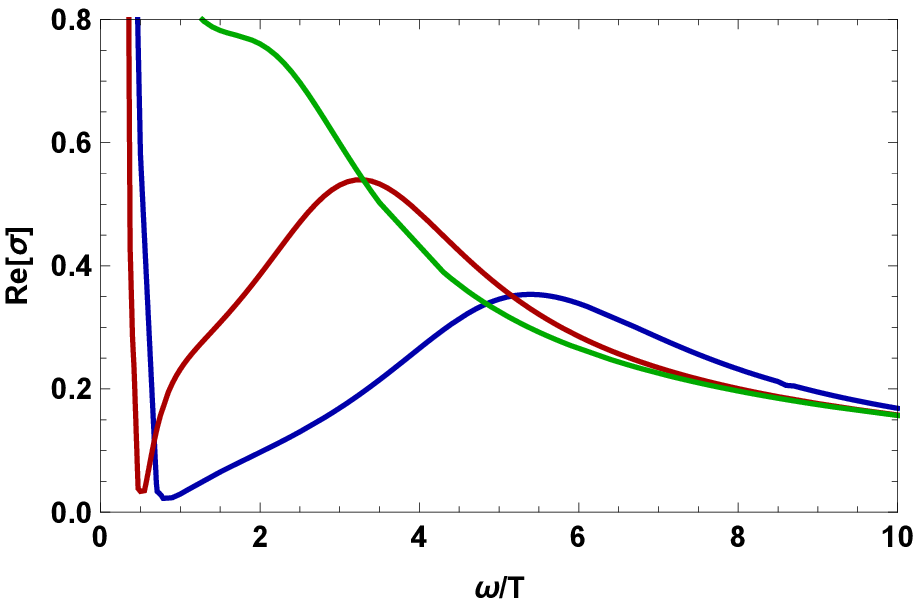}
\end{minipage}%
\begin{minipage}[b]{0.5\linewidth}
\includegraphics[width =2.5in,height=1.6in]{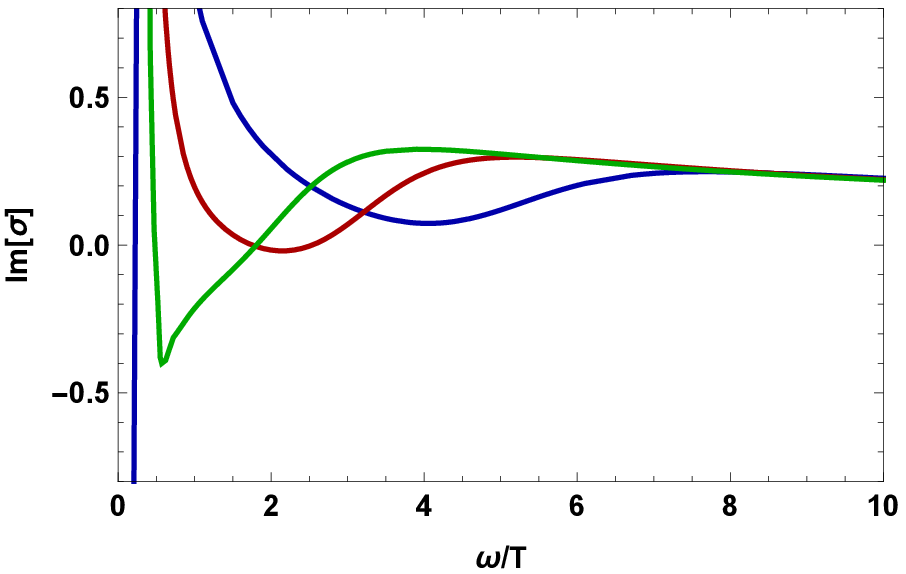}
\end{minipage}\quad
\caption{\label{fig:Conduct_0p3}For $m^2=1/4$, real and imaginary parts of the ac conductivity for superconducting phase are plotted with respect to $\omega/T$ for fixed value of $\gamma=0.3$. The blue, red and green curves correspond to the temperatures $T/T_c\approx0.4$ , $T/T_c\approx0.6$ and $T/T_c\approx0.8$ respectively.  } \end{figure}

At the boundary, perturbations have the following behavior
\begin{eqnarray}
&&A_x = - A \log(r) + B + \ldots,\;\;\; \rho_t = \frac{\rho_{t-} }{r^{\Delta_-}} + \frac{\rho_{t+} }{r^{\Delta_+}} + \ldots,\;\;\;\    \nonumber \\
&& \rho_t = \frac{\rho_{t-} }{r^{\Delta_-}} + \frac{\rho_{t+} }{r^{\Delta_+}} + \ldots,\;\;\;\Delta_{\pm} = \pm m.
\end{eqnarray}

Here, the leading term in the expansion of the perturbation $A_x$ is $-A \log(r)$, which leads us to treat $A$ as the source and $B$ as the response. Therefore, the Green's function is defined as \cite{Son:2002sd,Ren:2010ha}
\begin{equation}
G= -\frac{B}{A}=\lim_{r\rightarrow \infty}\frac{A_x+A_x^\prime r \log(r)}{r A_x^\prime},
\end{equation}
and with this form of perturbations, the conductivity becomes
\begin{equation}
\sigma_{xx}=-\frac{i}{\omega}G=\lim_{r\rightarrow \infty}\frac{i}{\omega}\frac{A_x+A_x^\prime r \log(r)}{r A_x^\prime}.\label{sigma}
\end{equation}

Non-renormalizable terms in the near boundary expansions of $\rho_t$ and $\rho_t^\dagger$ should be set to zero as a boundary condition and it can be  achieved in principle by shooting method. However, as discussed in \cite{Chen:2011ny} for a similar setup, the fact that the perturbation equations are linear in $A_x$, $\rho_t$ and $\rho_t^\dagger$ make this task considerably simpler. One can simply choose any set of ($c_1$, $c_2$, $c_3$) to generate a set of solutions for $A_x$, $\rho_t$ and $\rho_t^\dagger$, which do not have the correct behavior at the boundary. Generating three different sets of solutions ($A_x^{(i)}, \rho_t^{(i)}, \rho_t^{\dagger(i)}$) ($i=1,2,3$) and taking a linear combination of them in the following way
\begin{equation}
A_x=A_x^{(1)}+\alpha A_x^{(2)} + \beta A_x^{(3)}, \rho_t=\rho_t^{^\dagger(1)} +\alpha \rho_t^{(2)} + \beta \rho_t^{(3)}, \rho_t^\dagger=\rho_t^{^\dagger(1)} +\alpha \rho_t^{^\dagger(2)} + \beta \rho_t^{^\dagger(3)},  
\end{equation}
with $\alpha$ and $\beta$ being constants, produce another set of solutions since it is a linear system of equations. Furthermore, the constants ($\alpha$, $\beta$) can be chosen in such a way that the non-renormalizable modes of $\rho_t$ and $\rho_t^\dagger$ vanish and thus we can fix the physically consistent solution for $A_x$. Having the solution for $A_x$, the conductivity can be read off using the expression in eq.(\ref{sigma}).

In the figures (\ref{fig:Conduct_0p01}),(\ref{fig:Conduct_0p1}) and (\ref{fig:Conduct_0p3}) we present the numerical results for the ac conductivity of the (1+1)-dimensional boundary field theory and we show that the boundary theory exhibits the properties of a  {\it p-wave holographic superconductor}. In each of these figures, we demonstrate the comparative plots for the real part of the ac conductivity for different values of temperatures  $T/T_c\approx0.4$ , $T/T_c\approx0.6$ and $T/T_c\approx0.8$  at fixed values of the parameter $\gamma$. From these plots we observe that the real part of the ac conductivity diverges near $\omega/T=0$. The presence of a delta function can be verified from the divergent behavior of the imaginary part ($\text{Im}[\sigma]\sim\frac{1}{\omega}$) through the Kramers-Kronig relation. However, in some of the case that we consider, we do not observe such a behavior due to numerical limitations. There is also a characteristic peak which tends to be located at lower values of $\omega/T$  as the temperature increases. Moreover, from these figures it may be observed that there is a gap formation indicated by the significant decrease of the real part of ac conductivity at a certain frequency for lower value of the temperature  $T/T_c \approx0.4$ and higher value of the parameter $\gamma=0.3$. This shows that the superconducting phase is preferred for small values of temperature and higher values of the magnetic moment parameter $\gamma$.

\begin{figure}[H]
\centering
\begin{minipage}[b]{0.7\linewidth}
\includegraphics[width =3in,height=2.in]{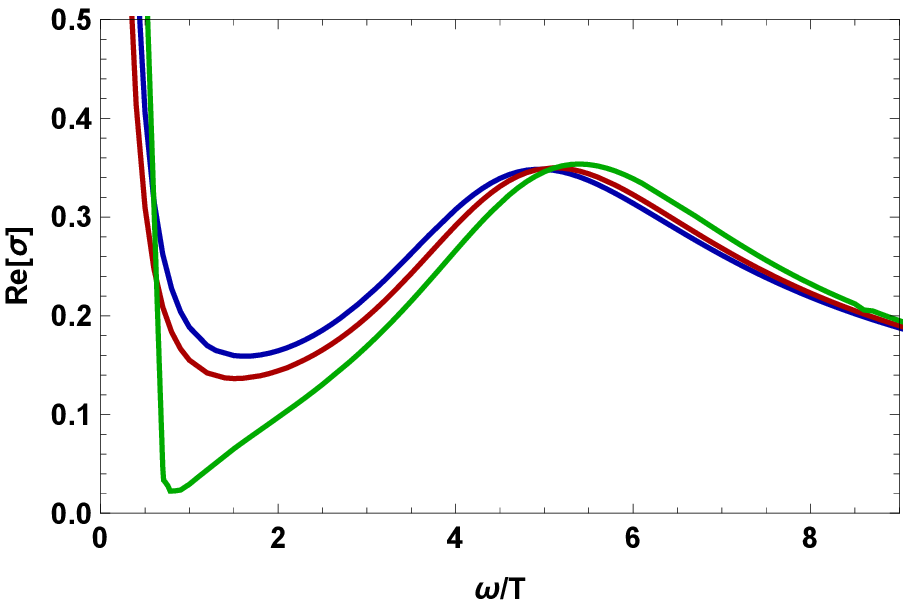}
\end{minipage}
\caption{\label{fig:ConductT_0p4} For $m^2=1/4$, real part of the ac conductivity for superconducting phase are plotted with respect to $\omega/T$ for fixed value of $T/T_c \approx0.4$. The blue, red and green curves correspond to the different values of the magnetic moment as $\gamma = 0.01$ , $\gamma = 0.1$ and $\gamma = 0.3$ respectively.}
 \end{figure} 

\begin{figure}[H]
\centering
\begin{minipage}[b]{0.7\linewidth}
\includegraphics[width =3in,height=2.in]{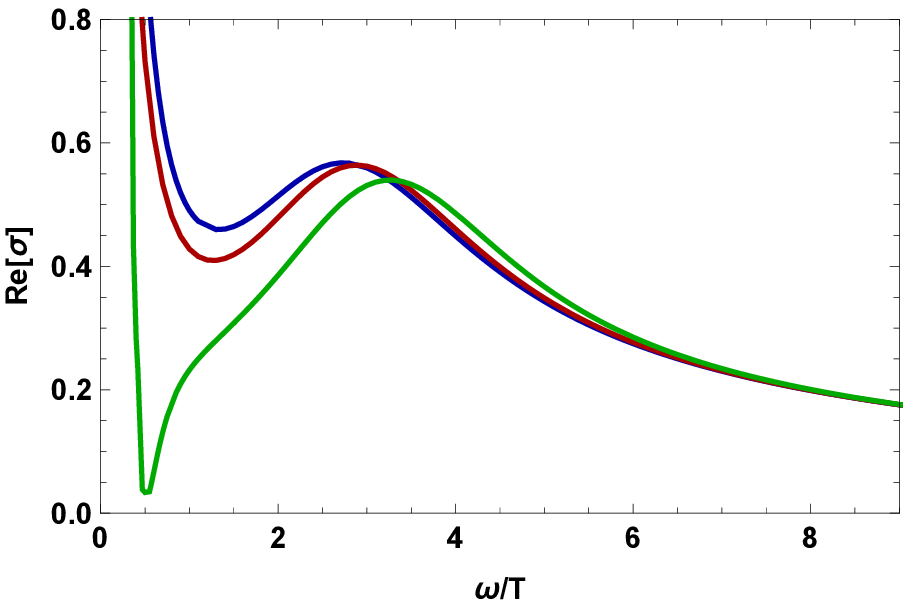}
\end{minipage}
\caption{\label{fig:ConductT_0p6}For $m^2=1/4$, real part of the ac conductivity for superconducting phase are plotted with respect to $\omega/T$ for fixed value of $T/T_c \approx0.6$. The blue, red and green curves correspond to the different values of the magnetic moment as $\gamma = 0.01$ , $\gamma = 0.1$ and $\gamma = 0.3$ respectively.}
 \end{figure} 

\begin{figure}[H]
\centering
\begin{minipage}[b]{0.7\linewidth}
\includegraphics[width =3in,height=2in]{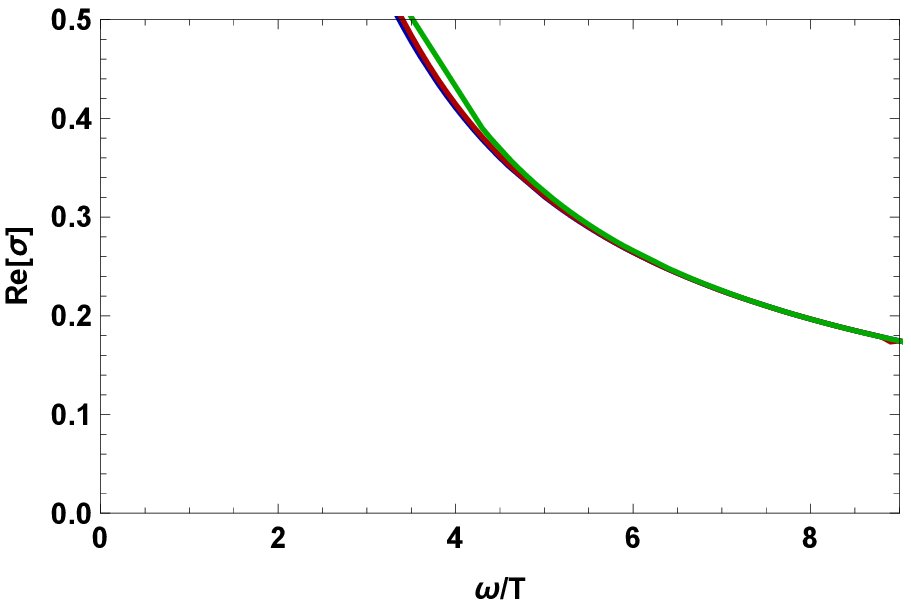}
\end{minipage}
\caption{\label{fig:ConductT_0p8}For $m^2=1/4$, real part of the ac conductivity for superconducting phase are plotted with respect to $\omega/T$ for fixed value of $T/T_c \approx0.8$. The blue, red and green curves correspond to the different values of the magnetic moment as $\gamma = 0.01$ , $\gamma = 0.1$ and $\gamma = 0.3$ respectively.}
\end{figure}

In the figures (\ref{fig:ConductT_0p4}),(\ref{fig:ConductT_0p6}),(\ref{fig:ConductT_0p8}), we have also shown the comparative plots  for the real part of the ac conductivity for different values of the parameter $\gamma=0.01,0.1$ and $0.3$ keeping the temperature fixed. From figure (\ref{fig:ConductT_0p8}), one sees that there is no significant effect of $\gamma$ at high temperatures. For the temperatures $T/T_c\approx0.4$ and $T/T_c\approx0.6$,  increasing the value of the parameter $\gamma$ shifts the location of the peak to the higher values of $\omega/T$. On the other hand, there is a characteristic dip which is shifted to the lower values of $\omega/T$ as $\gamma$ increases. We also see that it gets vanishingly deeper with increasing $\gamma$ which indicates a soft gap formation at higher value of the parameter $\gamma$.

\section{Conclusion}
In summary, here we have studied a model of (1+1)-dimensional p-wave holographic superconductors using Einstein-Maxwell theory coupled to a massive, charged vector field in three dimensions. The formation of vector hair was shown in the probe limit, which causes the condensation of the complex vector operator in the boundary field theory, breaking the U(1) and SO(1,1) symmetries spontaneously. This differs from the proposed gravitational duals of the (2+1)- dimensional p-wave holographic superconductors models studied earlier in \cite{Gubser:2008wv,Donos:2011ff,Donos:2012gg, Cai:2013pda,Cai:2013aca} as in these models the vector condensate breaks the both the U(1) and the rotational symmetry in the (2+1)-dimensional boundary. Computing the free energy of the boundary field theory corresponding to the (1+1)-dimensional p-wave holographic superconductor, we show that the superconducting phase is thermodynamically favorable than the normal phase which is also a common feature of earlier studied models of p-wave holographic superconductors.

We have also computed the conductivity of the boundary field theory in the superconducting phase and exhibit a feature of this model which is required by the consistency of perturbations introduced in the three dimensional bulk. The conductivity is affected by the magnetic moment term even in the absence of an external magnetic field. To understand the behavior of conductivity in a better way, we have plotted the $Re[\sigma]$ with respect to the ac frequency $\omega/T$ for different choices of the magnetic moment $\gamma$ and the temperature $T/T_c$ keeping one of them fixed at a time. For temperatures $T/T_c\approx0.4$ and $T/T_c\approx0.6$, before attaining an asymptotic for large values of $\omega/T$, $Re[\sigma]$ develops a dip as well as a peak at certain nonzero values of $\omega/T$. Increasing the magnetic moment $\gamma$ results in a sharper decrease to the dip and shifts it to the lower values of $\omega/T$.  On the contrary, when $\gamma$ takes higher values the peak is shifted to  the higher values of $\omega/T$. As we see from our analysis, the ac conductivity of the (1+1)-dimensional holographic p-wave superconductor is largely dependent on the values of the magnetic moment $\gamma$ and a soft gap formation occurs only for large values of $\gamma$ and low values of the temperature. This behavior is radically different from the higher dimensional case where the same model is employed to study ($2+1$)-dimensional p-wave holographic superconductors \cite{Cai:2013pda,Cai:2013aca}. In this case, there is no dependence of conductivity on the magnetic moment term in the absence of an external magnetic field. Additionally, one observes a hard gap formation which is characterized by the vanishing of the real part of ac conductivity up to a certain value of the frequency $\omega$. It would be very interesting to understand the interpretation of $\gamma$  and its influence on the ac conductivity from the boundary field theory perspective. 

There are number of directions one can pursue for further investigation of this model. The phase transition can be studied by taking the back-reaction into account to understand the nature of the condensate near $T=0$, where the probe limit approximation might fail to capture the details. One can also consider the duality on a more general background geometry such as Lifshitz black hole. One interesting possibility existing in three dimensional bulk is to consider a complex spin-1 field with topological mass as condensate and investigate the effect of parity breaking term on the conductivity. Analytical methods can be used to shed light on different aspects of this model in each of above mentioned scenarios. The last but not the least, it might be interesting to study the phase transition driven by an external magnetic field, which is a pseudo-scalar in three dimensions.

\section{Acknowledgements}

We thank Eric Bergshoeff for his guidance and supervision during the entire process. We would like to thank Arjun Bagchi, Souvik Banerjee, Yan Liu, Debaprasad Maity for useful discussions. We are also grateful to Mehmet Ozkan for discussions and his remarks on the manuscript. S.C. wishes to thank the Van Swinderen Institute for Particle Physics and Gravity and the Department of Physics, IIT Kanpur for warm hospitality during a major part of this work. G.A. acknowledges support by a grant of the Dutch Academy of Sciences (KNAW). S.C. is supported by Erasmus Mundus NAMASTE India-EU Grants as well as ISF/UGC joint research grant (UGC/PHY/2014236) during a part of this work. P.C. is supported by Grant No. 09/092(0846)/2012-EMR-I, from the Council of Scientific and Industrial Research (CSIR), India.


\begin{thebibliography}{10}

\bibitem{Maldacena:1997re}
Juan~Martin Maldacena.
\newblock {The Large N limit of superconformal field theories and
  supergravity}.
\newblock {\em Int.J.Theor.Phys.}, 38:1113--1133, 1999.

\bibitem{Gubser:1998bc}
S.S. Gubser, Igor~R. Klebanov, and Alexander~M. Polyakov.
\newblock {Gauge theory correlators from noncritical string theory}.
\newblock {\em Phys.Lett.}, B428:105--114, 1998.

\bibitem{Witten:1998qj}
Edward Witten.
\newblock {Anti-de Sitter space and holography}.
\newblock {\em Adv.Theor.Math.Phys.}, 2:253--291, 1998.

\bibitem{Aharony:1999ti}
Ofer Aharony, Steven~S. Gubser, Juan~Martin Maldacena, Hirosi Ooguri, and Yaron
  Oz.
\newblock {Large N field theories, string theory and gravity}.
\newblock {\em Phys.Rept.}, 323:183--386, 2000.

\bibitem{Witten:1998zw}
Edward Witten.
\newblock {Anti-de Sitter space, thermal phase transition, and confinement in
  gauge theories}.
\newblock {\em Adv.Theor.Math.Phys.}, 2:505--532, 1998.

\bibitem{Gubser:2008px}
Steven~S. Gubser.
\newblock {Breaking an Abelian gauge symmetry near a black hole horizon}.
\newblock {\em Phys.Rev.}, D78:065034, 2008.

\bibitem{Hartnoll:2008vx}
Sean~A. Hartnoll, Christopher~P. Herzog, and Gary~T. Horowitz.
\newblock {Building a Holographic Superconductor}.
\newblock {\em Phys.Rev.Lett.}, 101:031601, 2008.

\bibitem{Hartnoll:2008kx}
Sean~A. Hartnoll, Christopher~P. Herzog, and Gary~T. Horowitz.
\newblock {Holographic Superconductors}.
\newblock {\em JHEP}, 0812:015, 2008.

\bibitem{Horowitz:2008bn}
Gary~T. Horowitz and Matthew~M. Roberts.
\newblock {Holographic Superconductors with Various Condensates}.
\newblock {\em Phys.Rev.}, D78:126008, 2008.

\bibitem{Herzog:2009xv}
Christopher~P. Herzog.
\newblock {Lectures on Holographic Superfluidity and Superconductivity}.
\newblock {\em J.Phys.}, A42:343001, 2009.

\bibitem{Horowitz:2010gk}
Gary~T. Horowitz.
\newblock {Introduction to Holographic Superconductors}.
\newblock {\em Lect.Notes Phys.}, 828:313--347, 2011.

\bibitem{Herzog:2010vz}
Christopher~P. Herzog.
\newblock {An Analytic Holographic Superconductor}.
\newblock {\em Phys.Rev.}, D81:126009, 2010.

\bibitem{Sachdev:2010ch}
Subir Sachdev.
\newblock {Condensed Matter and AdS/CFT}.
\newblock {\em Lect.Notes Phys.}, 828:273--311, 2011.

\bibitem{Kim:2013oba}
Keun-Young Kim and Marika Taylor.
\newblock {Holographic d-wave superconductors}.
\newblock {\em JHEP}, 1308:112, 2013.

\bibitem{Benini:2010pr}
Francesco Benini, Christopher~P. Herzog, Rakibur Rahman, and Amos Yarom.
\newblock {Gauge gravity duality for d-wave superconductors: prospects and
  challenges}.
\newblock {\em JHEP}, 1011:137, 2010.

\bibitem{Chen:2010mk}
Jiunn-Wei Chen, Ying-Jer Kao, Debaprasad Maity, Wen-Yu Wen, and Chen-Pin Yeh.
\newblock {Towards A Holographic Model of D-Wave Superconductors}.
\newblock {\em Phys.Rev.}, D81:106008, 2010.

\bibitem{Gubser:2008wv}
Steven~S. Gubser and Silviu~S. Pufu.
\newblock {The Gravity dual of a p-wave superconductor}.
\newblock {\em JHEP}, 0811:033, 2008.

\bibitem{Donos:2011ff}
Aristomenis Donos and Jerome~P. Gauntlett.
\newblock {Holographic helical superconductors}.
\newblock {\em JHEP}, 12:091, 2011.

\bibitem{Donos:2012gg}
Aristomenis Donos and Jerome~P. Gauntlett.
\newblock {Helical superconducting black holes}.
\newblock {\em Phys. Rev. Lett.}, 108:211601, 2012.

\bibitem{Cai:2013pda}
Rong-Gen Cai, Song He, Li~Li, and Li-Fang Li.
\newblock {A Holographic Study on Vector Condensate Induced by a Magnetic
  Field}.
\newblock {\em JHEP}, 1312:036, 2013.

\bibitem{Cai:2013aca}
Rong-Gen Cai, Li~Li, and Li-Fang Li.
\newblock {A Holographic P-wave Superconductor Model}.
\newblock {\em JHEP}, 1401:032, 2014.

\bibitem{Gregory:2009fj}
Ruth Gregory, Sugumi Kanno, and Jiro Soda.
\newblock {Holographic Superconductors with Higher Curvature Corrections}.
\newblock {\em JHEP}, 10:010, 2009.

\bibitem{Liu:2010ka}
Yan Liu and Ya-Wen Sun.
\newblock {Holographic Superconductors from Einstein-Maxwell-Dilaton Gravity}.
\newblock {\em JHEP}, 07:099, 2010.

\bibitem{Gubser:2008pf}
Steven~S. Gubser and Abhinav Nellore.
\newblock {Low-temperature behavior of the Abelian Higgs model in anti-de
  Sitter space}.
\newblock {\em JHEP}, 04:008, 2009.

\bibitem{Horowitz:2010jq}
Gary~T. Horowitz and Benson Way.
\newblock {Complete Phase Diagrams for a Holographic Superconductor/Insulator
  System}.
\newblock {\em JHEP}, 1011:011, 2010.

\bibitem{Brihaye:2010mr}
Yves Brihaye and Betti Hartmann.
\newblock {Holographic Superconductors in 3+1 dimensions away from the probe
  limit}.
\newblock {\em Phys. Rev.}, D81:126008, 2010.

\bibitem{Domenech:2010nf}
Oriol Domenech, Marc Montull, Alex Pomarol, Alberto Salvio, and Pedro~J. Silva.
\newblock {Emergent Gauge Fields in Holographic Superconductors}.
\newblock {\em JHEP}, 08:033, 2010.

\bibitem{Roberts:2008ns}
Matthew~M. Roberts and Sean~A. Hartnoll.
\newblock {Pseudogap and time reversal breaking in a holographic
  superconductor}.
\newblock {\em JHEP}, 08:035, 2008.

\bibitem{Horowitz:2009ij}
Gary~T. Horowitz and Matthew~M. Roberts.
\newblock {Zero Temperature Limit of Holographic Superconductors}.
\newblock {\em JHEP}, 0911:015, 2009.

\bibitem{Amado:2009ts}
Irene Amado, Matthias Kaminski, and Karl Landsteiner.
\newblock {Hydrodynamics of Holographic Superconductors}.
\newblock {\em JHEP}, 05:021, 2009.

\bibitem{Nishioka:2009zj}
Tatsuma Nishioka, Shinsei Ryu, and Tadashi Takayanagi.
\newblock {Holographic Superconductor/Insulator Transition at Zero
  Temperature}.
\newblock {\em JHEP}, 1003:131, 2010.

\bibitem{Basu:2009vv}
Pallab Basu, Jianyang He, Anindya Mukherjee, and Hsien-Hang Shieh.
\newblock {Hard-gapped Holographic Superconductors}.
\newblock {\em Phys. Lett.}, B689:45--50, 2010.

\bibitem{Bu:2012zz}
Yan-Yan Bu.
\newblock {p-Wave superconductors in D-brane systems}.
\newblock {\em Nucl. Phys.}, B864:806--822, 2012.

\bibitem{Herzog:2009ci}
Christopher~P. Herzog and Silviu~S. Pufu.
\newblock {The Second Sound of SU(2)}.
\newblock {\em JHEP}, 04:126, 2009.

\bibitem{Ammon:2008fc}
Martin Ammon, Johanna Erdmenger, Matthias Kaminski, and Patrick Kerner.
\newblock {Superconductivity from gauge/gravity duality with flavor}.
\newblock {\em Phys. Lett.}, B680:516--520, 2009.

\bibitem{Ammon:2009fe}
Martin Ammon, Johanna Erdmenger, Matthias Kaminski, and Patrick Kerner.
\newblock {Flavor Superconductivity from Gauge/Gravity Duality}.
\newblock {\em JHEP}, 10:067, 2009.

\bibitem{Peeters:2009sr}
Kasper Peeters, Jonathan Powell, and Marija Zamaklar.
\newblock {Exploring colourful holographic superconductors}.
\newblock {\em JHEP}, 09:101, 2009.

\bibitem{Benini:2010qc}
Francesco Benini, Christopher~P. Herzog, and Amos Yarom.
\newblock {Holographic Fermi arcs and a d-wave gap}.
\newblock {\em Phys. Lett.}, B701:626--629, 2011.

\bibitem{Sonner:2009fk}
Julian Sonner.
\newblock {A Rotating Holographic Superconductor}.
\newblock {\em Phys. Rev.}, D80:084031, 2009.

\bibitem{Franco:2009yz}
Sebastian Franco, Antonio Garcia-Garcia, and Diego Rodriguez-Gomez.
\newblock {A General class of holographic superconductors}.
\newblock {\em JHEP}, 04:092, 2010.

\bibitem{Siopsis:2010uq}
George Siopsis and Jason Therrien.
\newblock {Analytic Calculation of Properties of Holographic Superconductors}.
\newblock {\em JHEP}, 05:013, 2010.

\bibitem{Ren:2010ha}
Jie Ren.
\newblock {One-dimensional holographic superconductor from AdS$_3$/CFT$_2$
  correspondence}.
\newblock {\em JHEP}, 11:055, 2010.

\bibitem{Liu:2011fy}
Yunqi Liu, Qiyuan Pan, and Bin Wang.
\newblock {Holographic superconductor developed in BTZ black hole background
  with backreactions}.
\newblock {\em Phys. Lett.}, B702:94--99, 2011.

\bibitem{Nurmagambetov:2011yt}
A.~J. Nurmagambetov.
\newblock {Analytical approach to phase transitions in rotating and
  non-rotating 2D holographic superconductors}.
\newblock 2011.

\bibitem{Zeng:2012xy}
Hua-Bi Zeng.
\newblock {One dimensional $s$-wave holographic superconductor with
  supercurrent}.
\newblock {\em Phys. Rev.}, D87(4):046009, 2013.

\bibitem{Momeni:2013waa}
Davood Momeni, Muhammad Raza, Mohammad~Reza Setare, and Ratbay Myrzakulov.
\newblock {Analytical Holographic Superconductor with Backreaction Using
  AdS3/CFT2}.
\newblock {\em Int. J. Theor. Phys.}, 52:2773--2783, 2013.

\bibitem{Momeni:2013eva}
Davood Momeni, Muhammad Raza, and Ratbay Myrzakulov.
\newblock {More on Superconductors via Gauge/Gravity Duality with Nonlinear
  Maxwell Field}.
\newblock {\em J. Grav.}, 2013:782512, 2013.

\bibitem{Chaturvedi:2013ova}
Pankaj Chaturvedi and Gautam Sengupta.
\newblock {Rotating BTZ Black Holes and One Dimensional Holographic
  Superconductors}.
\newblock {\em Phys. Rev.}, D90(4):046002, 2014.

\bibitem{Peng:2015lsa}
Yan Peng and Guohua Liu.
\newblock {General Holographic Superconductor Models in AdS BTZ Black Hole}.
\newblock {\em Int. J. Theor. Phys.}, 54(9):3424--3429, 2015.

\bibitem{Peng:2016mxd}
Yan Peng.
\newblock {Holographic insulator/superconductor transitions in the three
  dimensional AdS soliton}.
\newblock 2016.

\bibitem{Bu:2012qr}
Yanyan Bu.
\newblock {1+1-dimensional p-wave superconductors from intersecting D-branes}.
\newblock {\em Phys. Rev.}, D86:106005, 2012.

\bibitem{Breitenlohner:1982bm}
Peter Breitenlohner and Daniel~Z. Freedman.
\newblock {Positive Energy in anti-De Sitter Backgrounds and Gauged Extended
  Supergravity}.
\newblock {\em Phys. Lett.}, B115:197--201, 1982.

\bibitem{Hung:2009qk}
Ling-Yan Hung and Aninda Sinha.
\newblock {Holographic quantum liquids in 1+1 dimensions}.
\newblock {\em JHEP}, 01:114, 2010.

\bibitem{Chen:2011ny}
Jiunn-Wei Chen, Yu-Sheng Liu, and Debaprasad Maity.
\newblock {$d+id$ Holographic Superconductors}.
\newblock {\em JHEP}, 05:032, 2011.

\bibitem{Son:2002sd}
Dam~T. Son and Andrei~O. Starinets.
\newblock {Minkowski space correlators in AdS / CFT correspondence: Recipe and
  applications}.
\newblock {\em JHEP}, 09:042, 2002.

\end{thebibliography}
\end{document}